# Dimension reduction induced anisotropic magnetic thermal conductivity in hematite nanowire

Qing Xi,[1][*] Adili Ayiti,[1][*] Lan Dong,[3,4][†] Yuanyuan Wang,[3,4] Jun Zhou,[2][†] Xiangfan Xu[1][†]

[1]Center for Phononics and Thermal Energy Science, China-EU Joint Center for Nanophononics, School of Physics Science and Engineering, Tongji University, Shanghai 200092, China

[2]NNU-SULI Thermal Energy Research Center and Center for Quantum Transport and Thermal Energy Science, School of Physics and Technology, Nanjing Normal University, Nanjing 210023, China

[3] School of Energy and Materials, Shanghai Polytechnic University, Shanghai 201209, China

[4]Shanghai Engineering Research Center of Advanced Thermal Functional Materials, Shanghai Polytechnic University, Shanghai 201209, China

**ABSTRACT:** The thermophysical properties near the magnetic phase transition point is of great importance in the study of critical phenomenon. Low-dimensional materials are suggested to hold different thermophysical properties comparing to their bulk counterpart due to the dimension induced quantum confinement and anisotropy. In this work, we measured the thermal conductivity of α-$Fe_2O_3$ nanowires along [110] direction (growing direction) with temperature from 100K to 150K and found a dip of thermal conductivity near the Morin temperature. We found the thermal conductivity near Morin temperature varies with the angle between magnetic field and [110] direction of nanowire. More specifically, an angular-dependent thermal conductivity is observed, due to the magnetic field induced movement of magnetic domain wall. The angle corresponding to the maximum of thermal conductivity varies near the Morin transition temperature, due to the different magnetic easy axis as suggested by our calculation based on magnetic anisotropy energy. This angular dependence of thermal conductivity indicates that the easy axis of α-$Fe_2O_3$ nanowires is different from bulk α-$Fe_2O_3$ due to the geometric anisotropy.

* These authors contribute equally to this work and should be considered co-first authors

†Correspondence: donglan@sspu.edu.cn；zhoujunzhou@njnu.edu.cn；xuxiangfan@tongji.edu.cn；

## INTRODUCTION

The magnetic properties of hematite ($\alpha$-Fe$_2$O$_3$) have been extensively studied both theoretically and experimentally[1-3]. Its sublattice magnetization is determined by the overall effect of the Zeeman term of external field, the Heisenberg exchange interaction, the Dzialoshinskii exchange interaction and the crystalline anisotropic free energy for a rhombohedral structure [2, 4, 5]. As a result, bulk hematite could experience a magnetic phase transition, from an antiferromagnet, with magnetization along the c-axis, to a canted weak ferromagnet, with magnetization in the basal plane and a slight canting away from the antiferromagnetic axis. It is known as Morin transition, which is a first-order spin-reorientation transition and the corresponding transition temperature is named as Morin transition temperature ($T_M$). Below the Morin transition temperature, a large enough magnetic field will also induce a similar phase transition, which is called spin flopping and the critical filed for spin flopping is $H_{AF}$[2]. For bulk hematite, $T_M \cong$ 263 K[2] and $H_{AF} \cong$ 6 T[6]. Since the deflection of moments might alter the phonon and magnon modes and influence the magnon-phonon interaction at the same time, thermal conductivity at the vicinity of Morin transition is of great interest, and has been investigated in antiferromagnets such as FeCl$_2$[7], CoBr$_2$.6H$_2$O[8], MnCl$_2$.4H$_2$O[9], Nd$_2$CuO$_4$[10] and Pr$_{1.3}$La$_{0.7}$CuO$_4$[11]. Thermal conductivity of $\alpha$-Fe$_2$O$_3$, especially $\alpha$-Fe$_2$O$_3$ nanostructures, are suggested to possess many interesting properties corresponding to its magnetic properties. Moreover, the dimension reduction induced quantum confinement and anisotropy is also interesting.

Many novel magnetic properties have been found in nanoscale hematite materials like nanoparticles and nanowires. For example, a reduced transition temperature has been observed in many measurements on hematite nanoparticles[12] and nanowires[5, 13, 14], where the minimum $T_M$ reported to be reduced to 80 K, depending on the diameter of hematite nanowires/nanoparticles. Lu *et al.* provided a thermodynamic analytic model to describe the size dependence of $T_M$ [15]. Moreover, a core-shell structure might exist in nanostructured $\alpha$-Fe$_2$O$_3$, which makes the real structure of $\alpha$-Fe$_2$O$_3$ nanowires more complicated[16, 17]. Chionel *et al.* observed a Morin transition at 123K for hematite nanowires (diameter=100-200nm, length=10μm), and the observation of coercive field supported their assumption that an antiferromagnetic core was surrounded by a ferrimagnetic shell below $T_M$ [17]. It has been reported that the phonon properties are sensitive to the magnetic order, such as phonon dispersion, anharmonicity, and thermal conductivity[18]. Therefore, nanoscale $\alpha$-Fe$_2$O$_3$ might exhibit some unique thermal transport behaviors. Wang *et al.* has measured thermal conductivity of $\alpha$-Fe$_2$O$_3$ nanowires in a wide temperature region from 20K to 300K[19], and they observed that phonon is the dominant heat carriers in $\alpha$-Fe$_2$O$_3$. However, thermal conductivity at the vicinity of Morin temperature still lacks systematical study. Since the magnetic spin orientation has been reported to be affected by the nanostructure[4, 5, 12, 13], it is also interesting whether thermal conductivity is sensitive to the spin orientation and magnetic domain structures.

In this letter, we investigate the temperature effect and magnetic field effect on the thermal conductivity of hematite ($\alpha$-Fe$_2$O$_3$) nanowires along the growing direction, i.e. [110]-direction. The Morin transition temperature of $\alpha$-Fe$_2$O$_3$ nanowires in this experiment is around 127K, in accordance with previous literatures. An anomalous reduction of thermal conductivity has been observed at the vicinity of Morin temperature. We further conduct a systematic investigation of the thermal conductivity of $\alpha$-Fe$_2$O$_3$ nanowire as a function of the direction of external magnetic field. The magnetic field dependence of thermal conductivity reveals that the magnetic anisotropic energy of $\alpha$-Fe$_2$O$_3$ nanowires is determined by both the magnetocrytalline anisotropy and geometric anisotropy, which is different from bulk $\alpha$-Fe$_2$O$_3$.

**EXPERIMENTAL AND SAMPLE CHARACTERIZATION**

Figure 1(a) presents the morphology of a single $\alpha$-Fe$_2$O$_3$ nanowire which is suspended on a MEMS (Micro-Electro-Mechanical System) device, suitable for thermal conductivity measurement. Individual $\alpha$-Fe$_2$O$_3$ nanowires were picked up by nanomanipulators with tungsten needles and transferred to the suspended MEMS device, which was operated under an optical microscope system (Olympus SZ6045). In order to reduce the thermal contact resistance between the nanowire and the platinum electrodes of MEMS device, the electron beam induced Pt/C deposition (EBID) was used to fix the two ends of the $\alpha$-Fe$_2$O$_3$ nanowire on the electrodes [20]. In our experiments we made four Pt/C fixed areas to optimize the thermal contact resistance. The hysteresis loops of $\alpha$-Fe$_2$O$_3$ nanowires in 300K and 10K are shown in Fig. 1(b). The magnetization saturates above 3000 Oe both below and above the Morin transition temperature. Figure 1 (c) shows the temperature dependence of the field-cooling (FC) and zero field-cooling (ZFC) magnetization of the $\alpha$-Fe$_2$O$_3$ nanowires. The abrupt change of magnetization (*dM/dT*) at 127K represents the Morin transition temperature of $\alpha$-Fe$_2$O$_3$ nanowires. Below 127K, $\alpha$-Fe$_2$O$_3$ nanowire is in an antiferromagnetic phase and the magnetization is relatively small. Above 127K, $\alpha$-Fe$_2$O$_3$ nanowire is in a canted weak ferromagnetic phase, thus the magnetization is relatively large. The X-ray scattering patterns of $\alpha$-Fe$_2$O$_3$ nanowire is given in Fig. 1(d). The [110] peak indicates that the nanowire is grown with the [110]-direction, confirmed by the High Resolution Transmission Electron Microscope image[19].

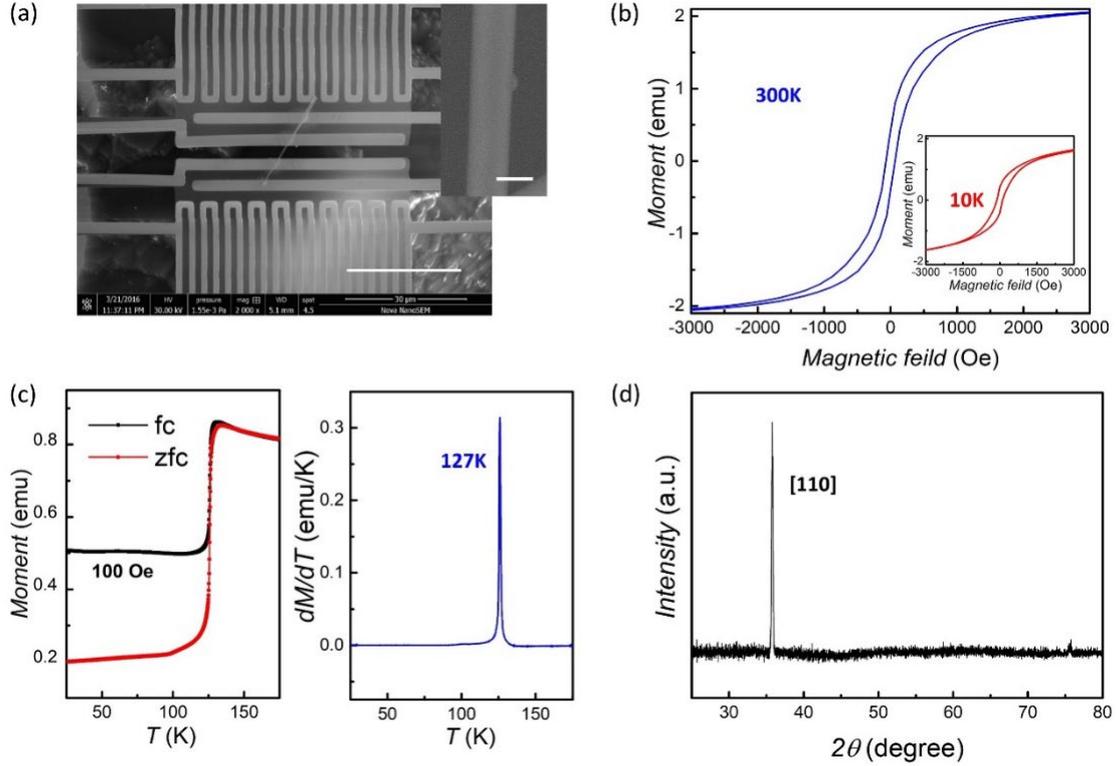

Fig. 1 (a) Scanning Electron Microscope (SEM) images of a single α-$Fe_2O_3$ nanowire on the suspended MEMS device, with 3.3μm in length and 583 nm in diameter; (b) Hysteresis loops of α-$Fe_2O_3$ nanowires at 300K and 10K, exhibiting the magnetic saturation of α-$Fe_2O_3$ nanowire to above $H$ = 3000 Oe. (c) Field-cooling (FC) and zero field-cooling (ZFC) magnetization of the α-$Fe_2O_3$ nanowires under $H$ = 100 Oe. The Morin temperature of the α-$Fe_2O_3$ nanowires is determined to be around 127 K from $dM/dT$; (d) X-ray diffraction (XRD) patterns of α-$Fe_2O_3$ nanowires. It reveals that the growing direction of α-$Fe_2O_3$ nanowire is [110]-direction.

In order to systematically investigate the change of thermal conductivity of α-$Fe_2O_3$ nanostructure induced by the 'spin-flop', we use the traditional thermal bridge method to measure the thermal conductivity of a suspended single α-$Fe_2O_3$ nanowire around the Morin temperature [21-24]. The suspended MEMS device with a single α-$Fe_2O_3$ nanowire after EBID process was placed into the cryogenic system with magnet (Oxford, TeslatronPT) that could provide ±12T magnetic field. The vacuum circumstance of this system could reach $10^{-4}$ Pa to optimize the effects of thermal convection and thermal radiation during thermal conduction measurements. We have characterized the thermal conductivity ($\kappa$) along [110]-direction of α-$Fe_2O_3$ nanowires from 20K to 300K which are given in our previous work Ref. [19]. It is observed that $\kappa$ increases with temperature at low temperatures and decrease with temperature at high temperatures, which is a typical behavior of thermal conductivity of crystals. It reveals that phonons are dominant heat carriers in α-$Fe_2O_3$ nanowires. Phonon-boundary scattering is the dominant scattering processes at low temperature and phonon-phonon scattering dominates at high temperatures.

## RESULTS AND DISCUSSION

The thermal conductivity of the α-Fe$_2$O$_3$ nanowire near the Morin temperature was measured as shown in Fig. 2(a). Considering the α-Fe$_2$O$_3$ nanowire is sensitive to magnetic field near Morin transition temperature (~127K), here we applied 1T magnetic field in the process of thermal transport measurement and found an anomalous dip of thermal conductivity near 127K. The main heat carriers in α-Fe$_2$O$_3$ are phonons and the mean free path of phonons is dominated by the phonon-boundary scattering, the phonon-phonon scattering and the phonon-magnon scattering. Therefore, the observed dip of thermal conductivity at Morin temperature has two possible origins. One is due to the discontinuous change of the magnon dispersion and the change of magnon-phonon scattering processes at the phase transition[9]. The other is the change of magnetic domain walls at the Morin transition[25]. In previous literatures, it was suggested that the interaction between magnons and phonons are weak in hematite[26], because the velocity of magnons is larger than phonons. The presence of the domain walls and their motions are more likely to influence the elastic properties of hematite[27]. In addition, we also found that the thermal conductivity of α-Fe$_2$O$_3$ nanowire is sensitive to the angle $\theta$ (shown in Fig.2(b)), where $\theta$ is the angle between the y axis and magnetic field. When $\theta$ is 90° and the magnetic field remains at 1T, the extent of dip in thermal conductivity was further increased (shown in Fig.2(a) with solid blue circles). In order to clarify the dominant scattering mechanism of phonons, we take a systematic investigation on the angle dependence of thermal conductivity.

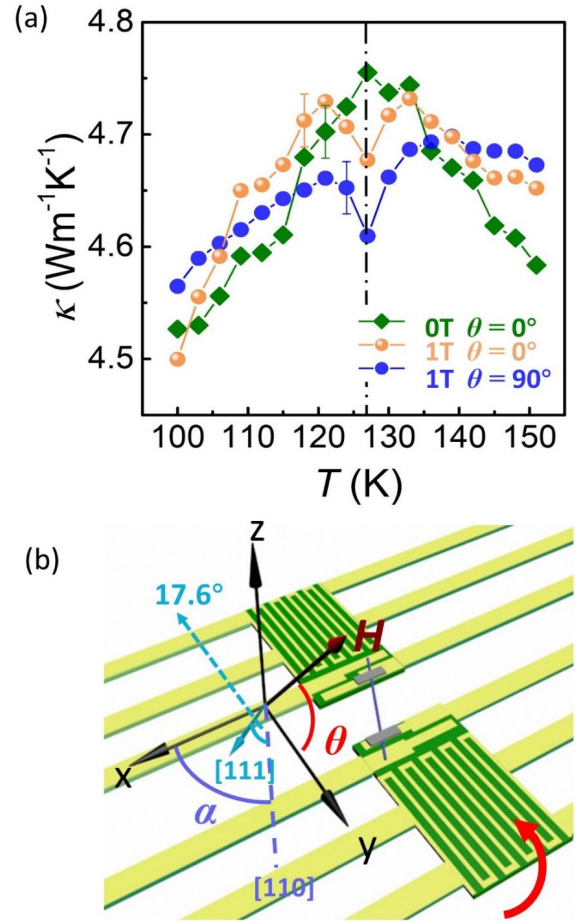

Fig. 2(a)Thermal conductivity of α-Fe$_2$O$_3$ nanowire (3.3μm in length and 583nm in diameter). Green diamonds represent thermal conductivity under zero field, orange circles and solid blue circles represent thermal conductivity of nanowire under 1T magnetic field with $\theta$ equals 0°and 90°, black dash dots exhibit Morin transition temperature at 127K; (b) Definition of angles in a spherical polar coordinate system with z axis perpendicular to the experimental platform. Magnetic field is rotated within zy plane, and $\theta$ is the angle between y axis and magnetic field. α is the angle between nanowire axis and x direction.

The thermal conductivity of α-Fe$_2$O$_3$ nanowire was further measured in detail when the platform of Magnet Cryogenic System plane rotates, as shown in xy plane in Fig.2(b). The rotation of the platform essentially changes the angle $\theta$. To ensure that the magnetic field is large enough to have an observable effect on the reorientation of magnetization, we keep the magnetic field at 1T. Fig. 3(a) and 3(c) show thermal conductivity of α-Fe$_2$O$_3$ nanowire as a function of angle $\theta$ at 100K (below Morin transition temperature 127K) and 150K (above Morin transition temperature), respectively. Fig. 3(b) shows thermal conductivity of α-Fe$_2$O$_3$ nanowire as a function of angle at Morin transition temperature under different strength of magnetic field. The results under different strength of magnetic field have also been separately plotted in Fig. 3(d-f). At 100K, $\kappa$ increases from 4.48 Wm$^{-1}$K$^{-1}$ to 4.55Wm$^{-1}$K$^{-1}$ with $\theta$ from 0° to 30°, decreases from 4.55Wm$^{-1}$K$^{-1}$to 4.48 Wm$^{-1}$K$^{-1}$ with $\theta$ from 30° to 60°, and increases again to 4.55Wm$^-$

$^{-1}K^{-1}$ with $\theta$ from 60° to 90°. The variation of thermal conductivity is around 1.5%. Since the platform of Magnet Cryogenic System could only rotate 90°, we then inversely changed the external magnetic field and measured the thermal conductivity with $\theta$ from 180° to 270°. The measured thermal conductivity is in good rotational symmetry with the data from 0° to 90°. At 150K, $\kappa$ increases from 4.70 $Wm^{-1}K^{-1}$ to 4.74 $Wm^{-1}K^{-1}$ with $\theta$ from 0° to 45°, and decrease again to 4.70 $Wm^{-1}K^{-1}$ with $\theta$ from 45° to 90°. After reversing the direction of magnetic field, the measured thermal conductivity is also in good rotational symmetry with the data from 0° to 90°. At higher temperatures, the variation of thermal conductivity is relatively small, only 0.8%. In general, we found that below the Morin transition temperature, $\theta$ dependence of $\kappa$ has its maximum at 30°, 90°, 210°, and 270°. Above Morin transition temperature, $\theta$ dependence of $\kappa$ has its maximum at 45° and 225°. At the Morin transition temperature, thermal conductivity shows two different angle dependence when changing the strength of magnetic field. At 0.01T, $\kappa$ has its maximum at 0°, 90°, 180° and 270°, while at 0.1T and 1T, $\kappa$ has its maximum at 45° and 225°, which is similar to the case of 150K. The observed $\theta$ dependence of $\kappa$ is ascribed to the magnetic field induced domain wall based on the interplay between magnetic field and magnetic anisotropic energy of α-$Fe_2O_3$ nanowire as discussed below.

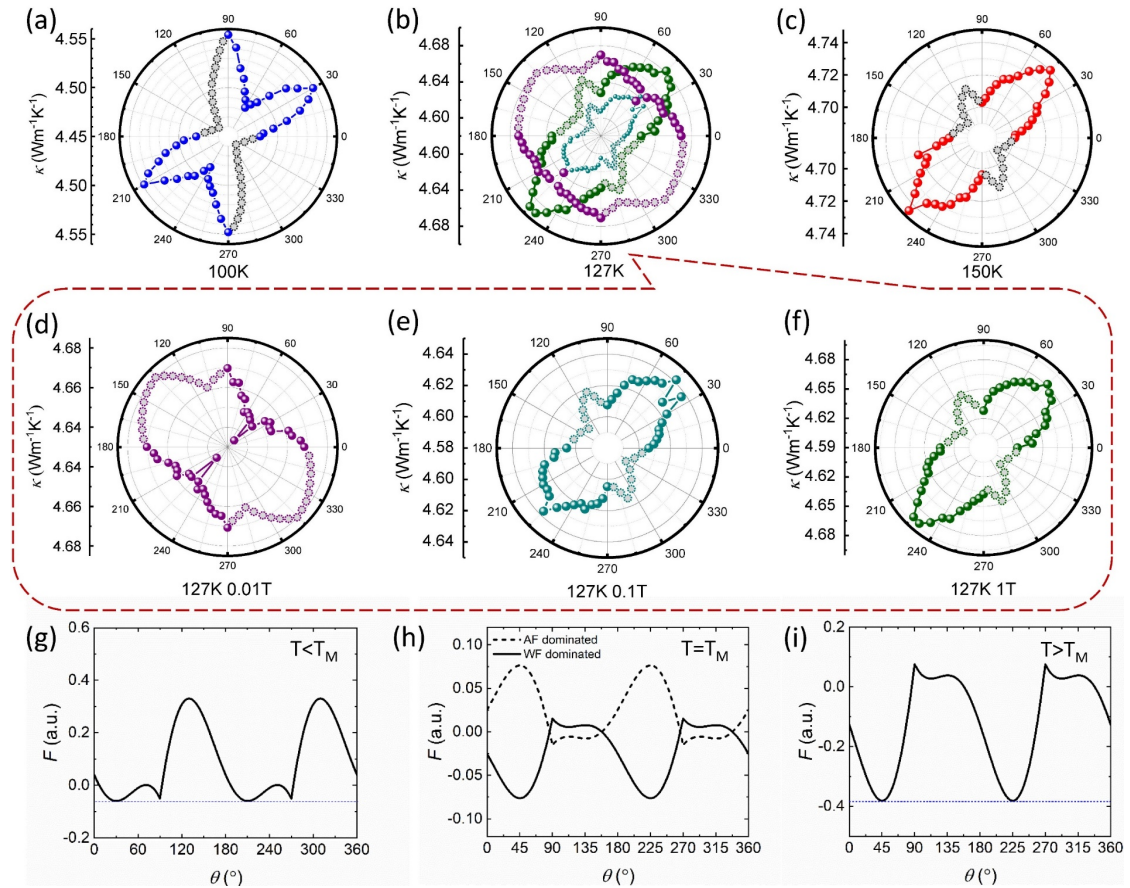

Fig. 3 Thermal conductivity of α-$Fe_2O_3$ nanowire as a function of $\theta$ at 100K (a), 127K (b) and 150K (c). Solid symbols are measured data and open symbols are plotted as a prediction of the thermal conductivity according to the $\theta$ dependence of magnetic energy $F$ as guides for the eyes. Symbols with different color in (b) correspond to data measured under different magnetic field, which are separately plotted in (d-f). (g-i) are magnetic energy $F$ as a function of $\theta$ according to Eq. (3) with $K_s=K_1$. The solid line and dash line in (h) represent $F$ of the states when canted weak ferromagnetic (WF) phase dominates or antiferromagnetic (AF) phase dominates at Morin transition temperature, respectively.

In bulk samples, the magnetic anisotropic energy is mainly determined by the magnetocrystalline anisotropy. While in nanostructures, the geometric anisotropy is non-negligible and the magnetic anisotropic energy becomes sensitive to geometric structures [12, 13, 28-30]. For example, Gee et al. reported that the sublattice magnetization oriented 28° with respect to c-axis ([111]-direction) at antiferromagnetic phase in 40-nm-sized spherical hematite particles, other than along the c-axis as in bulk hematite[12]. In ferromagnetic nanowires, such as Fe, Co and Ni, magnetization tends to lie along the long axis of nanowires[31, 32]. However, Kim et al. observed that the easy axis of hematite nanowires is perpendicular to the nanowire axis [13]. Here we give a brief discussion on the magnetic anisotropic energy of α-Fe$_2$O$_3$ nanowires considering both magnetocrystalline anisotropy and geometric anisotropy.

From a phenomenological point of view, the magnetocrystalline anisotropy energy of bulk hematite is expressed to first order as [2, 4, 5]

$$F_U = -\frac{K_1}{2}(\cos^2\theta_1 + \cos^2\theta_2), \qquad (1)$$

Where $K_1$ is magnetocrystalline anisotropic energy constant, $\theta_1$ and $\theta_2$ are angle between magnetization of sublattice and [111] direction. A change in sign of $K_1$ explains the Morin transition, namely, $K_1 > 0$ when $T < T_M$, and $K_1 < 0$ when $T > T_M$. It has been verified that $\theta_1 = 0$ and $\theta_2 = \pi$ when $T < T_M$ and $\theta_1 = \theta_2 = \frac{\pi}{2}$ when $T > T_M$ [2, 4, 5]. It means that below Morin transition temperature, the magnetic moments lie along the [111] direction. While above Morin transition temperature, the magnetic moments lie in the basal plane perpendicular to [111]. In nanowires, the geometric anisotropic energy is expressed as

$$F_s = K_s|\cos\theta_N|, \qquad (2)$$

where $K_s$ is effective geometric anisotropic energy constant, and $\theta_N$ is the angle between magnetization and nanowire axis ([110]-direction). Here we assume $K_S > 0$ when $T < T_M$, and $K_S < 0$ when $T > T_M$, which is deduced from current experimental measurements and requires further validation[13]. The total anisotropic energy is the sum of magnetocrystalline anisotropic energy and geometric anisotropic energy, namely, $F = F_U + F_s$.

Based on the experiment platform of our measurement, the direction perpendicular to the experiment platform could be defined as z axis of the spherical coordination. The rotating plane of platform is defined as yz plane (as shown in Fig. 2(b)). Since nanowire is not exactly placed in the rotational plane of experimental platform, there is an angle between nanowire and x axis, which is labeled as $\alpha$. In our measurement, $\alpha \approx 45°$ as shown in Fig. 1(a). In this coordination, the total anisotropic energy in yz plane as a function of $\theta$ is expressed as

$$F = F_U + F_s = -K_1[\sin\theta_c\sin\phi_c\cos\theta + \cos\theta_c\sin\theta]^2 + K_s|\sin\alpha\cos\theta|. \quad (3)$$

where $\theta_c$ and $\phi_c$ are polar angle and azimuthal angle of c-axis ([111] direction) of nanowire, respectively. The relation $\cos\theta_1 = \cos\theta_2 = \sin\theta_c\sin\phi_c\cos\theta + \cos\theta_c\sin\theta$ and $\cos\theta_N = \cos(\frac{\pi}{2} - \alpha)\cos\theta$ has been applied here. Since it is hard to determine the direction of [111] in the spherical coordination, we could only fit the results under the restriction that the angle between [110] and [111] is 17.6°. Here we approximately take $K_s = K_1$, and the $\theta$ dependence of $F$ is plotted in Fig. 3(g-i) according to Eq. (3). As shown in Fig. 3(g), the minimum of total anisotropic energy appears at $\theta = 30°$, $90°$, $210°$ and $270°$ at $T < T_M$ with $\theta_c = 77°$. As shown in Fig. 3(i), the minimum appears at $\theta = 45°$ and $225°$ at $T > T_M$ with $\theta_c = 106°$. A more rigorous consideration must include the anisotropic energy in all directions, but here we only considered the angle dependence of anisotropic energy in yz plane for simplicity, because the effect of anisotropic energy in other directions tends to be similar for magnetic field rotated in yz plane.

Then we discuss the effect of magnetic field on thermal conductivity based on the landscape of magnetic anisotropic energy. If the external magnetic field is parallel to the direction corresponding to the minimum of anisotropic energy, the spin will orient along the magnetic field, because it leads to the lowest anisotropic energy of system. In this case, the magnetic domain along the direction of magnetic field will grow and domain walls will be reduced [25]. However, if the magnetic field is along a direction other than the minimum anisotropic energy direction, the spin tends to orient along the direction corresponding to a local minimum of the anisotropic energy. Since the configuration of each domain will only reach its local minimum of anisotropic energy, there will remain a lot of domains with magnetization of different directions [25]. As a result, thermal conductivity is higher when magnetic field is aligned along a direction corresponding to a lower anisotropic energy. A maximum thermal conductivity is expected at the angle corresponding to the minimum anisotropic energy. It explains well the $\theta$ dependence of thermal conductivity of α-Fe$_2$O$_3$ nanowire as shown in Fig. 3(a) and (c). The situation is more complicated at the Morin transition temperature, as the two phases coexist and the $\theta$ dependence is dominated by the dominant phase. At 0.01T, there are more antiferromagnetic (AF) phase in the nanowire, so κ has its maximum at 0°, 90°, 180° and 270°, agreeing with the $\theta$ dependence of $F$ for AF phase dominated structure with $\theta_c = 106°$ (plotted by dotted line in Fig. 3(h)). At 0.1T and 1T, there are more canted weak ferromagnetic (WF) phase in the nanowire, and κ has its maximum at 45° and 225°, agreeing with the $\theta$ dependence of $F$ for WF phase dominated structure with $\theta_c = 106°$ (plotted by solid line in Fig. 3(h)). We also theoretically predicted the thermal conductivity at angles which has not been measured in our experimental platform according to the $\theta$ dependence of $F$ (open circles in Fig. 3(a-f)). Although the domain wall effect based on the interplay between magnetic field and magnetic anisotropic energy explains well the result, it remains an open question whether magnon-phonon interaction also plays a role for the $\theta$ dependence of thermal conductivity.

## CONCLUSION

In summary, we measured the thermal conductivity of α-Fe$_2$O$_3$ nanowires along [110]-direction from 100K to 150K. Thermal conductivity as a function of the angle $\theta$ between [110]-direction and magnetic field has been systematically investigated. It has been observed that the Morin temperature of our α-Fe$_2$O$_3$ nanowire sample is around 127K, and thermal conductivity shows a dip structure at the vicinity of Morin temperature under external magnetic field, which has been attributed to the interaction between domain walls and phonons at the transition point. Another interesting phenomenon is that thermal conductivity is sensitive to the direction of magnetic field. When external magnetic field is along the direction corresponding to the minimum of anisotropic energy, thermal conductivity will increase as the domain walls are reduced. The dependence of thermal conductivity on the direction of magnetic field reveals that the anisotropic energy of α-Fe$_2$O$_3$ nanowires is determined by both the magnetocrytalline anisotropy and geometric anisotropy. Our experimental platform might shed new light on studying the magnetic order-dependent phonon properties, and further theoretical and experimental investigation are recommended to unveiling the underlying mechanisms.

## METHOD

*Traditional thermal bridge method*

The suspended MEMS device contains a single α-Fe$_2$O$_3$ nanowirewas placed into Magnet Cryogenic System (Oxford, TeslatronPT) with a high vacuum in the order of 1×10$^{-4}$ Pa to reduce the thermal convection[33]. The two Pt/SiNx resistive thermometers served as heater and sensor on the MEMS device, were used to characterize the temperature rise ($\Delta T_h$ and $\Delta T_s$) of the both ends of α-Fe$_2$O$_3$ nanowire. The combined current within 1μA AC current and 70μA DC current (Current source, Keithley 6221) were added to the heater resistor, the DC current provides heat power and the AC current was used to measure the change of resistance of heater. Part of the heating current created on the heater flows through the single nanowire and increases the sensor temperature, the other part is directed to the circumstances through the supporting SiNx beams. A same AC current was added to the sensor which was used to measure the change of resistance of sensor[34]. The thermal conductance of the supporting SiNx beams and α-Fe2O3nanowire could be obtained by: $G_b = \frac{Q_{tot}}{\Delta T_h + \Delta T_s}$ and $G_s = \frac{G_b \Delta T_s}{\Delta T_h - \Delta T_s}$, where $Q_{tot}$ is the total heat flow adds on the heater, $\Delta T_h$ and $\Delta T_s$ act as the temperature rise of the heater and sensor respectively. $G_s$ represents the total thermal conductance of α-Fe2O3nanowire, and $G_b$ is the thermal conductance of the supporting SiNx beams. The thermal conductivity of the α-Fe2O3nanowire can be obtained by: $\kappa = G_s \frac{L_s}{A}$, where $\kappa$ is the thermal conductivity of the α-Fe2O3nanowire, $L_s$ and $A$ are the length and cross section area of the α-Fe2O3nanowire. Here we consider the cross-sectional area of the nanowire to be circular and $A = \pi d^2/4$.

The measurement accuracy of the thermal conductance of α-Fe2O3 nanowire is directly related to the temperature measurement accuracy of sensor thermometer. Thanks to the high vacuum environment and temperature control time of no less than 5 hours, we obtained the extremely high temperature measurement accuracy ($\Delta T_{TMA} \sim 5mK$). The measurement accuracy of the thermal conductance ($G_{SA} = \frac{G_b \Delta T_{TMA}}{\Delta T_h - \Delta T_s}$) can therefore be estimated in the order of $10^{-11}$ W/K. This is at least four orders of magnitude lower than the thermal conductivity of the α-Fe2O3nanowire.

The uncertainty in the thermal conductivity of α-Fe2O3nanowire is estimated using formula as follow:

$$\frac{\delta \kappa}{\kappa} = \sqrt{(\frac{\delta G}{G})^2 + (2\frac{\delta d}{d})^2}$$

Considering the diameter uncertainty of α-Fe2O3 nanowire, we give 0.5% measurement error in our thermal conductivity measurement.

*Measurement of angle dependence thermal conductivity*

The platform of the sample holder of Magnet Cryogenic System (Oxford, Teslatron PT) could rotate from 0° to 90° and the direction of the magnetic field could be reversed. It is reasonable that the angle $\theta$ dependence of thermal conductivity could be characterized at 0° to 90° and 180° to 270° (magnetic field reverses). $\theta$ is the angle between the nanowire growing direction ([110]-direction) and external magnetic field. Here we choose external magnetic field around 1T to achieve the magnetic saturation in α-Fe2O3nanowire. The thermal conductivity measurement at different angles $\theta$ is still based on the thermal bridge method. In order to achieve the high-precision measurement of thermal conductivity in our experiments, we stayed for more than 5 hours at each angle to wait for the system conditions reaching uniformity.

**ACKNOWLEDGEMENT**

The work was supported by the National Natural Science Foundation of China (No. 11890703 & 11935010 & 12174286 & 12004242), by the Key-Area Research and Development Program of Guangdong Province (No. 2020B010190004), by the Open Fund of Zhejiang Provincial Key Laboratory of Quantum Technology and Device (No. 20190301), and by Shanghai Rising-Star Program (No. 21QA1403300).